\documentclass[runningheads]{llncs}

\usepackage{algorithm, algpseudocode, color, colortbl, graphicx, tcolorbox}
\usepackage{url, array, cite, multirow, mathtools, framed, amsfonts}
\usepackage{amsmath, amssymb, caption, newfloat}
\usepackage{diagbox, booktabs, subcaption, appendix, alltt}
\usepackage{listings, mathrsfs, xcolor, afterpage, bm, amsbsy}
\usepackage[multiple]{footmisc}
\usepackage{xspace}
\usepackage[cal=boondox]{mathalfa}
\usepackage{soul}
\DeclareMathAlphabet{\mathcal}{OMS}{cmsy}{m}{n}
{}
\makeatletter
\newcommand{\@chapapp}{\relax}%
\makeatother

\DeclareMathAlphabet\mathbfcal{OMS}{cmsy}{b}{n}

\newcommand{\ignore}[1]{}

\definecolor{bittersweet}{rgb}{1.0, 0.44, 0.37}
\newcommand\ghada[1]{\nbc{GA}{#1}{bittersweet}}

\usepackage{ifthen}
\usepackage[normalem]{ulem} 
\usepackage{xcolor}
\usepackage{amssymb}

\newboolean{showedits}
\setboolean{showedits}{true} 
\ifthenelse{\boolean{showedits}}
{
	\newcommand{\del}[1]{\textcolor{red}{\sout{#1}}} 
}{
	\newcommand{\del}[1]{} 
	
}

\newboolean{showcomments}
\setboolean{showcomments}{true}
\newcommand{\id}[1]{$-$Id: scgPaper.tex 32478 2010-04-29 09:11:32Z oscar $-$}

\ifthenelse{\boolean{showcomments}}
{\newcommand{\nbc}[3]{
 {\colorbox{#3}{\bfseries\sffamily\scriptsize\textcolor{white}{#1}}}
 {\textcolor{#3}{\sf\small$\blacktriangleright$\textit{#2}$\blacktriangleleft$}}}
 }
{\newcommand{\nbc}[3]{}
 \renewcommand{\del}[1]{} 
 }

\definecolor{jccolor}{rgb}{0.2,0.4,0.6}
\definecolor{clcolour}{rgb}{0.5,0.7,0.9}

\usepackage{wasysym}

\begin{document}
%
\title{Bet and Attack: Incentive Compatible Collaborative Attacks Using Smart Contracts}
\titlerunning{Bet and Attack}
\author{Zahra Motaqy\inst{1}\thanks{Most work done while at University of Tehran.} \and
Ghada Almashaqbeh\inst{2} \and
Behnam Bahrak\inst{3} \and
Naser Yazdani\inst{4}}
\authorrunning{Z. Motaqy et al.}
%
%
\institute{University of Connecticut,
\email{raha@uconn.edu}\\ 
\and 
University of Connecticut,
\email{ghada.almashaqbeh@uconn.edu} \\
\and 
University of Tehran,
\email{bahrak@ut.ac.ir} \\
\and University of Tehran,
\email{yazdani@ut.ac.ir}}

\maketitle

\begin{abstract}
Smart contract-enabled blockchains allow building decentralized applications in which mutually-distrusted parties can work together. Recently, oracle services emerged to provide these applications with real-world data feeds. Unfortunately, these capabilities have been used for malicious purposes under what is called criminal smart contracts. A few works explored this dark side and showed a variety of such attacks. However, none of them considered collaborative attacks against targets that reside outside the blockchain ecosystem.

In this paper, we bridge this gap and introduce a smart contract-based framework that allows a sponsor to orchestrate a collaborative attack among (pseudo)anonymous attackers and reward them for that. While all previous works required a technique to quantify an attacker's individual contribution, which could be infeasible with respect to real-world targets, our framework avoids that. This is done by developing a novel scheme for trustless collaboration through betting. That is, attackers bet on an event (i.e., the attack takes place) and then work on making that event happen (i.e., perform the attack). By taking DDoS as a usecase, we formulate attackers' interaction as a game, and formally prove that these attackers will collaborate in proportion to the amount of their bets in the game's unique equilibrium. We also model our framework and its reward function as an incentive mechanism and prove that it is a strategy-proof and budget-balanced one. Finally, we conduct numerical simulations to demonstrate the equilibrium behavior of our framework.
\end{abstract}

\vspace{-10pt}

\section{Introduction}
\label{intro}
\vspace{-4pt}
Cryptocurrencies and blockchain technology continue to build innovative computing models and economic tools that can reshape the services and systems around us. Fueled by the huge interest this technology received, researchers and practitioners alike are racing to build new applications and improve existing ones. Smart contracts facilitate this process; individuals can deploy arbitrary code on a blockchain, allowing for trustless collaboration between participants under terms enforced by the contract execution. More recently, and to achieve the quest of allowing blockchain to react to real-world events, the concept of oracles has been introduced~\cite{provable,chainlink,aeternity}. These are external services that supplement a smart contract with information about specific real-world events.

However, these new capabilities have been used for malicious purposes as well. This falls under what is called criminal smart contracts (CSCs), a term that was first coined by Juels et al.~\cite{juels2016ring}. They showed that CSCs can be used for cryptographic key theft, leakage of confidential information, and real-world crimes such as murder. Since then, several studies have proposed new smart contract-based attacks mainly related to bribery to disrupt the mining process~\cite{velner2017smart,mccorry2018smart,chen2018game,judmayer2019pay}. Nonetheless, these works were limited to (collaborative) attacks on the cryptocurrency/blockchain itself rather than outsider real-world targets. Juels et al.\cite{juels2016ring} considered real-world targets, but only for solo attackers.

Collaborative attacks are known to be more devastating~\cite{vogt2007army, xu2008collaborative,vogt2006attack,enisa, computerweekly, DarkReading, nazario2009politically}. This raises the question of whether CSCs can facilitate such attacks in the real world. Addressing this question is challenging; (pseudo)anonymity of blockchain users leads to incomplete information about the attackers. Also, dealing with attackers in the real world makes it hard to measure and verify their contribution. Attackers are represented by random-looking public keys in a CSC who tend to hide their involvement in a given attack to avoid legal consequences. This is different from attacks with targets within the blockchain ecosystem itself, where usually cryptographic primitives comes for the rescue to prove a claimed  contribution~\cite{judmayer2019pay}. This is even easier in the context of a solo attacker; the fact that an attack took place at the promised time/location implies that the work has been done~\cite{juels2016ring}. Verifying contribution is essential to ensure that self-interested parties cannot collect their rewards without doing the required work. \\

\vspace{-4pt}
\noindent{\bf Contributions.} In this paper, we bridge this gap by showing how smart contracts can be used to perform collaborative attacks against real-world targets. We develop an incentive compatible framework that allows a sponsor to orchestrate a collaborative attack among (pseudo)anonymous attackers and reward them for that. Our framework mitigates the requirement of quantifying attackers' individual contributions by introducing a betting-based technique to allow trustless collaboration. In particular, attackers bet on an event (i.e., the attack takes place) and then work on making that event happen (i.e., perform the attack). By designing an incentive-compatible reward mechanism, these attackers will be motivated to deliver the work represented by the value of their bets, and hence, obtain their rewards. To the best of our knowledge, our work is the first generic CSC-based collaborative attack framework against real-work targets.

As a use case, we consider distributed denial of service (DDoS) attacks.\footnote{We note that~\cite{rodrigues2020sc} dealt with smart contract-based DDoS, but the work is very high level and lacks many important details, making it hard to assess its feasibility.} We design a smart contract to perform this attack, and we formulate attackers' interaction as a game. Then we formally prove that these attackers will collaborate in proportion to the amount of their bets in the game's unique equilibrium. We also model our framework and its reward function as an incentive mechanism and prove that it is strategy-proof (i.e., attackers will not misrepresent the amounts of their bets) and that it is budget-balanced (i.e., the total rewards allocated to the attackers doesn't exceed the deposited attack rewards). Through numerical simulations, we study the impact of several parameters, including the reward function, total amount of bets, and attack cost, on the attack outcome. We show that in a typical scenario, the proposed incentive mechanism provides individual rationality and fairness for the collaborating attackers.

We argue that showing feasibility of collaborative attacks using CSCs is essential to identifying such threats, and devising secure countermeasures. Moreover, and although our focus is on attacks in this paper, we believe that our framework can facilitate benign collaboration between users. For example, it can be employed in blockchain-based decentralized systems offering digital services, such as content delivery~\cite{noia} and file storage~\cite{filecoin}, to incentivize peers to act honestly while serving others. Such a feature has a huge impact on system efficiency. That is, incentive compatibility mechanisms (when applicable) can be used to defend against potential threats instead of (usually computationally-heavy) cryptographic mechanisms. \\


\vspace{-4pt}
\noindent{\bf Related work.}
Many incidents were reported on cybercriminals collaboration in the context of malware and massive DDoS attacks~\cite{enisa, computerweekly, DarkReading, nazario2009politically}. Attackers trade goods, services, and money through the phishing marketplace and even advertise their demands~\cite{abad2005economy}. Moreover, botnets can be even rented for spam campaigns and DDoS attacks~\cite{networkworld}.

Usually these attacks involve some trust assumption that a sponsor will reward attackers for participation. CSCs can replace such assumption by providing an automated and transparent way for coordination and sponsorship~\cite{juels2016ring,brunoni2017smart,o2017smart}. As mentioned before, this observation was first investigated by Juel et al.~\cite{juels2016ring}. They proved how the CSCs they developed can be commission fair, meaning that neither the sponsor nor the attackers can cheat. Another work that considered real world targets is~\cite{delgado2020blockchain} who proposed a semi-autonomous (file) ransomware architecture using CSCs. Their scheme allows paying for individual files or reimbursing the victim if decryption is invalid. Despite handling real-world targets, these works focused only on solo attackers.

Another line of work on CSCs focused on bribery to attack the cryptocurrency or blockchain itself. \cite{velner2017smart} showed an attack to allow a mining pool manager to destroy competing mining pools, and~\cite{mccorry2018smart,judmayer2019pay} showed how a sponsor can bribe miners to pursue a mining strategy that benefits him (e.g., to allow him to revert a transaction or to double spend). CSC-based attacks on mining have been systematically analyzed in~\cite{judmayer2020sok} under the umbrella of algorithmic incentive manipulation attacks. On the game theoretic front, Chen et al.~\cite{chen2018game} introduced a game modeling of a bribery contract. They showed that in any Nash equilibrium, a sponsor cannot win the majority of the votes unless he/she controls more than 20\% of the total bribing budget. Although the aforementioned bribery attacks consider collaborative attackers, their target is the blockchain/cryptocurrency itself. Also, these works implement techniques to enable attackers to provide cryptographic proofs that the desired mining strategy has been performed. As we mentioned before, such aspect is hard to achieve when the target resides outside the blockchain ecosystem. Our work handles these issues by showing the feasibility of CSC-based collaborative attacks against real-world targets without measuring the individual contribution of attackers.

\section{A Model for a CSC-based Collaborative Attack}
\label{ddos-attack}
In this section, we present the blockchain and threat models we adopt in this work, along with a security notion for CSCs. After that, we introduce the proposed CSC-based collaborative attack model with DDoS attacks as a usecase.

\vspace{-4pt}
\subsection{Blockchain Model}
\vspace{-4pt}
We deal with permissionless public blockchains that support smart contracts, such as Ethereum~\cite{wood2014ethereum}. Hence, the code of the smart contract, its state, and all messages (or transactions) sent to this contract are logged in the clear on the blockchain. Any party needs an account in order to interact with a smart contract, referred to as externally owned account (EOA) in Ethereum. This account is identified by the public key of its owner, and has a public state that is mainly concerned with the account currency balance. Anyone can create any number of EOAs and participate in a CSC, and no real identities are required. Although we consider Ethereum in our CSC model, any other smart contract infrastructure can be used given that its scripting language can represent the CSC functionality.

We require the blockchain to have access to oracles that provide authenticated real-world data feeds. A popular example of such oracle services is Provable~\cite{provable}. In the context of a CSC, the attack sponsor (who is also the CSC creator) will specify the metrics that the CSC will query from the oracle that will quantify the outcome of the attack. For example, in our DDoS usecase, we use the response delay as the attack outcome metric, where excessive delays means an overwhelmed (or a down) server. Provable can measure this metric by sending http requests to the target server and report the response to the CSC for which parameters like delay can be measured.

\vspace{-4pt}
\subsection{Threat Model}
\vspace{-4pt}
We adopt the following threat model in this paper:
\begin{itemize}
\vspace{-5 pt}
\itemsep0em
\item The blockchain is secure in the sense that the majority of the mining power is honest. So the confirmed state of the blockchain contains only valid transactions, and that any attempt to rewrite the blockchain will fail with overwhelming probability. We also assume liveness, meaning that messages cannot be blocked or delayed beyond some bounded duration, and availability in the sense that the blockchain records are accessible at anytime. 

\item The oracle service is secure (i.e., provide valid digitally signed data), always available, and capable of measuring the attack result in terms of some predefined metrics.\footnote{It is the responsibility of the attack sponsor to pick metrics for which there is a secure oracle service that cam measure and report them.}

\item No trusted party of any type exists. Also, attackers involved in any CSC are mutually-distrusted and rational so they will act based on what maximizes their individual profits. Moreover, these attackers are regarded as independent risk-neutral decision-makers; they will be interested in maximizing the expected value of their utility.\footnote{While it is true that attackers are not always risk-neutral, we assume that is the case here for simplicity. For an analysis of non-risk neutral attackers, see \cite{qian2015robust}.}
\end{itemize}

A \emph{fully successful attack} in our model is defined as meeting the desired attack result set by the CSC sponsor. The difference between the actual and desired attack result denote the success level of an attack. It should be noted that in some attacks such success level is not applicable; it is either the attack has been done or not (e.g., steal a cryptographic key).

For security, we adopt the notions of \emph{correctness} and \emph{commission-fairness} proposed in~\cite{juels2016ring}. A correct CSC is one that implements the attack protocol as designed by the sponsor. And a commission-fair CSC is one that guarantees that neither the sponsor nor the perpetrator of the crime can cheat; attack sponsor cannot avoid paying attackers for the work they have done, and attackers cannot collect rewards for work they have not done. Note that \emph{work} here refers to pursuing the attack as per the sponsor's desired metrics.

\vspace{-4pt}
\subsection{Attack Model}
\begin{figure}[t!]
\centering
\includegraphics[height= 1.3in, width = 0.6\textwidth]{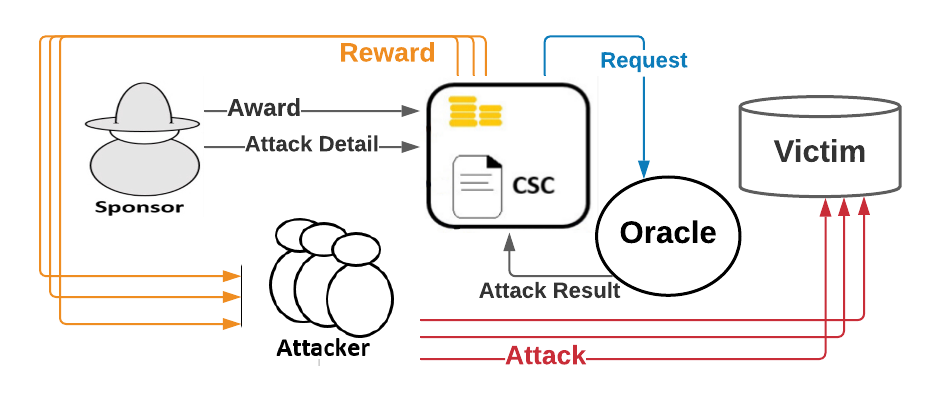}
\vspace{-6pt}
\caption{A model for a CSC-based collaborative attack.}
\label{coll-attack-model}
\vspace{-10pt}
\end{figure}

\begin{figure}[t!]
\centering
\includegraphics[width=0.9\textwidth,height=1.8in]{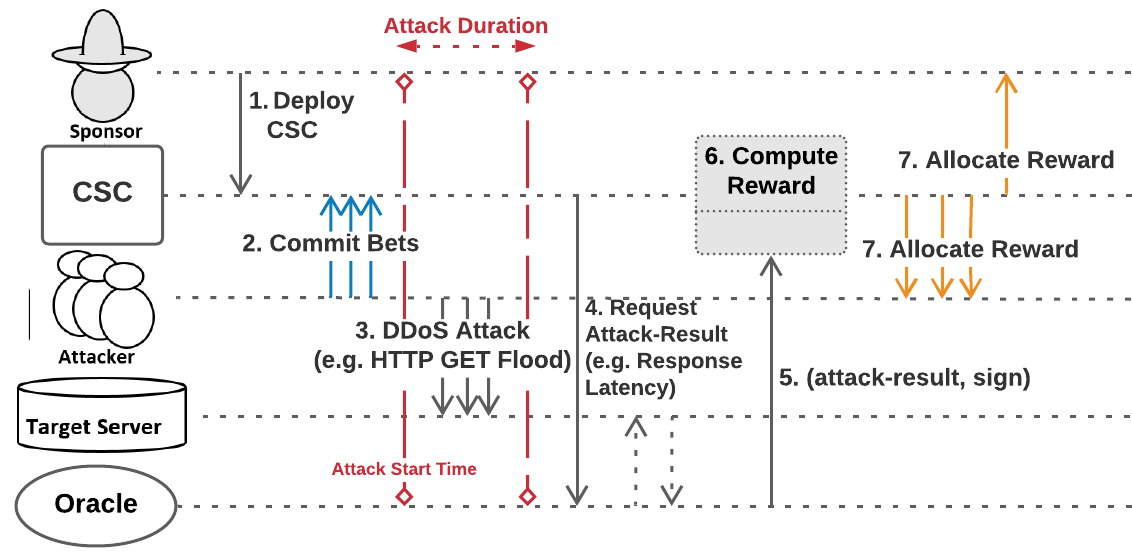}
\vspace{-6pt}
\caption{Sequence diagram for a DDoS attack using CSC}
\label{Fig12}
\vspace{-12pt}
\end{figure}

\vspace{-4pt}
At a high level, and as shown in Figure~\ref{coll-attack-model}, a CSC-based collaborative attack is composed of three phases: a CSC design and deployment phase, an attack phase, and a reward allocation phase. In order to make the discussion easier to follow, we present these phases with a running example. In particular, we consider one of the devastating cybersecurity attacks, namely, DDoS attacks. The sequence diagram of this attack is captured by Figure~\ref{Fig12}, while the following paragraphs elaborate on the attack protocol according to our framework phases. \\

\vspace{-6pt}
\noindent{\bf Phase 1. Design and deployment of CSC.} In this phase, an attack sponsor designs the CSC functionality. This includes defining all APIs and methods needed to orchestrate the attack and distribute the rewards. This in addition to attack parameters such as attack target, duration (measured in rounds, where a round is the time needed to mine a block on the blockchain), the metrics used to measure the attack result, and the reward allocation function. For DDoS, the target can be a specific server, and the metric that we use is response delay. The oracle service Provable can measure such quantity by simply sending HTTP requests to the specified target and report the response back to the CSC for which parameters like delay can be measured~\cite{provable}.

The sponsor then creates a smart contract implementing the CSC functionality and publishes it on the Ethereum blockchain. He will also deposit an award for the attackers in the CSC account. Once published, and since the contract code is public on the blockchain, interested attackers can evaluate the terms and assess if it is feasible for them to participate in the attack. For example, they check their availability during the attack duration and whether they have the resources required to achieve the attack goal. If feasible, each attacker will submit a bet to the CSC's account during the betting period. This period starts when the contract is confirmed on the blockchain until the attack starting time.

In terms of DDoS, an attacker checks if he can reaches the target server, and that he can afford sending traffic to overwhelm the server. In this context, this attacker can, for example, rent botnets to achieve that~\cite{networkworld}. Based on the amount of traffic this attacker can afford, which represents his amount of contribution in the attack, the attacker will choose his bet value and deposit in the DDoS CSC. \\

\vspace{-6pt}
\noindent{\bf Phase 2. The attack.} The second phase of our framework is the attack phase. Once the deposits are made by both the sponsor and the attackers, and the attack period starts, these attackers will launch the attack. Each attacker may choose any strategy given that it achieves the attack goal set by the sponsor. For DDoS, an attacker sends appropriate traffic to overwhelm the target.\footnote{While it is common that the target defends itself by identifying and filtering the attack traffic, for simplicity we assume that attackers generate effective attack traffic (traffic that passes the defense walls and gets to the target server).} Since all attackers are sending traffic during this period, and although they do not know and do not trust each other, they are collaborating against the same target. CSC allows this automated collaboration coordination without placing trust in anyone. The attack phase continue until the end of the attack period specified in the contract. \\

\vspace{-6pt}
\noindent{\bf Phase 3. Reward allocation.} In the third phase, the CSC queries the oracle to obtain the attack result. This can be done by having the sponsor send a transaction to invoke a function in CSC that sends a request to the oracle. The oracle then fetches the attack result and return it by executing a callback function in CSC. Based on the attack goal, this can happen either when the specified attack duration is over (e.g., check that the target server is down) or during that period (i.e., check that the server response delay is long enough). If multiple measures are reported, then the average, minimum or maximum can be computed (recall this is part of the terms that the sponsor specifies).

After that, the CSC computes the total currency value of the sponsor award and the attackers' bets. Then, it distributes this amount among the sponsor and the attackers based on the reward allocation function and the attack result. If the attack is fully successful, the total currency will be distributed among the attackers. If the attack is unsuccessful, then this currency will go to the sponsor (i.e., attackers are punished by taking away their bets). For attacks where there is a success level some where in between, as in DDoS, both attackers and sponsors will get part of the total currency. The distribution of this amount will be based on a reward allocation function that the sponsor chooses. In Sections~\ref{game-model} and~\ref{inc-comp} we thoroughly discuss the details of the reward allocation mechanism.

At the end, the CSC sends payment transactions to the sponsor and/or attackers' accounts (these accounts are defined by the attackers public keys used when registering the bets). 


We note that despite advances in devising countermeasures against DDoS~\cite{ddos-defense}, these attacks cannot be fully prevented. Nonetheless, it is important to point out that given that the CSC is public on the blockchain, the target could be aware of the attack (assuming he is inspecting the blockchain regularly). Hence, the target can  employ defense strategies during the specified attack period. This will increase the cost of performing the attack as we will show in Section~\ref{analysis}.\\

\begin{figure}[t!]
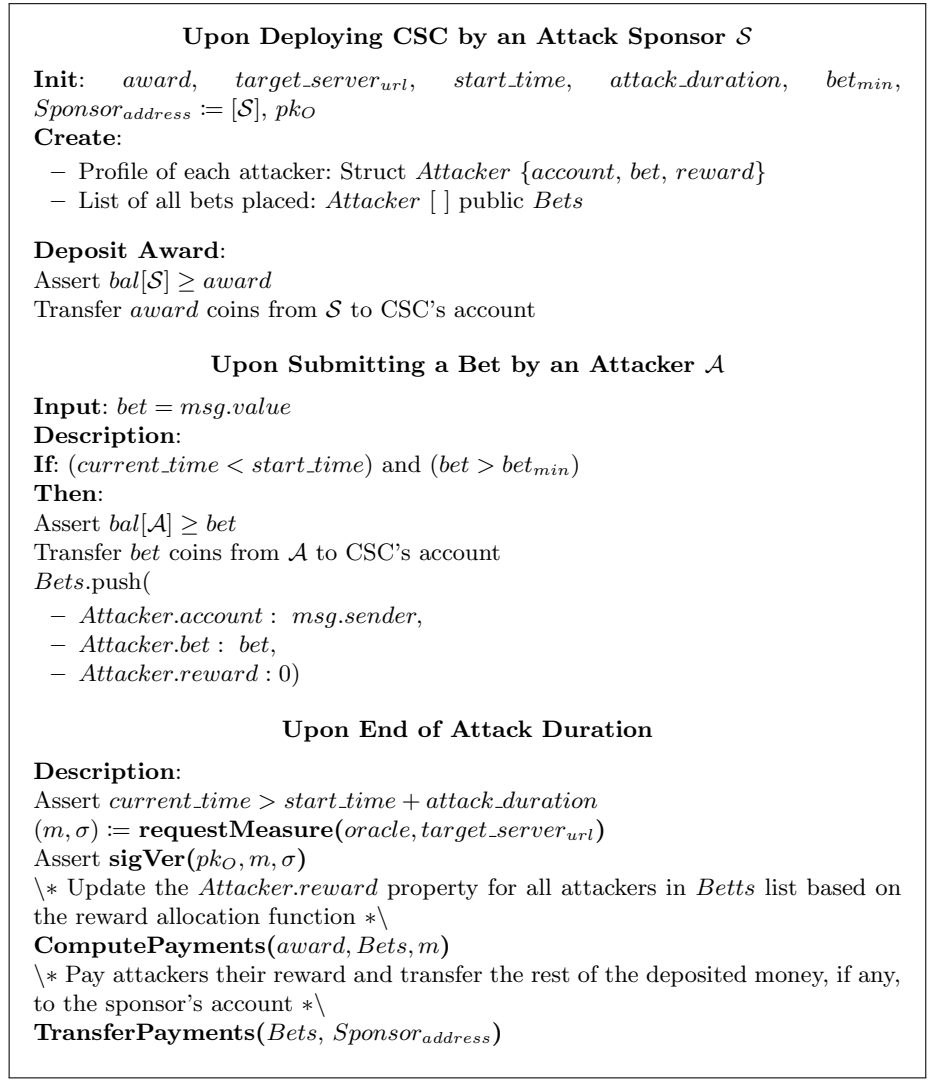

\begin{framed}
    \begin{center}
    {\bf Upon Deploying CSC by an Attack Sponsor $\mathcal{S}$}\\
    \end{center}
    \vspace{-5pt}
    {\bf Init}: $award$, $target\_server_{url}$, $start\_time$, $attack\_duration$, $bet_{min}$, $Sponsor_{address} \coloneqq [\mathcal{S}] $, $pk_O$\\
    {\bf Create}:
    \begin{itemize}
    \itemsep0em
    \vspace{-6 pt}
        \item Profile of each attacker: Struct $Attacker$ \{$account$, $bet$, $reward$\}
        \item List of all bets placed: $Attacker$ [ ] public $Bets$
    \end{itemize}
    {\bf Deposit Award}:\\
    Assert $bal[\mathcal{S}] \geq award$\\
    Transfer $award$ coins from $\mathcal{S}$ to CSC's account

\begin{center}
    {\bf Upon Submitting a Bet by an Attacker $\mathcal{A}$}
\end{center}
\vspace{-5pt}
{\bf Input}: $bet = msg.value$\\
{\bf Description}: \\
    {\bf If}:
    ($current\_time < start\_time $) and ($ bet > bet_{min}$)\\
    {\bf Then}:\\
    Assert $bal[\mathcal{A}] \geq bet$\\
    Transfer $bet$ coins from $\mathcal{A}$ to CSC's account\\
    $Bets$.push(
    \begin{itemize}
    \itemsep0em
    \vspace{-6 pt}
        \item $ Attacker.account: \ msg.sender, $
        \item $ Attacker.bet: \ bet, $
        \item $ Attacker.reward: 0 )$
    \end{itemize}

\begin{center}
    \textbf{Upon End of Attack Duration}
\end{center}
\vspace{-5pt}
{\bf Description}:

    Assert $current\_time > start\_time + attack\_duration$
    
	$(m, \sigma) \coloneqq $ \textbf{requestMeasure(}$oracle, target\_server_{url}$\textbf{)}\\
	Assert \textbf{sigVer(}$pk_{O}, m, \sigma$\textbf{)} \\
	$\backslash\ast$ Update the $Attacker.reward$ property for all attackers in $Betts$ list based on the reward allocation function $\ast\backslash$
	
    \textbf{ComputePayments(}$award, Bets, m$\textbf{)}
    
    $\backslash\ast$ Pay attackers their reward and transfer the rest of the deposited money, if any, to the sponsor's account $\ast\backslash$
    
	\textbf{TransferPayments(}$Bets$, $Sponsor_{address}$\textbf{)}\\
	
\vspace{-6pt}
\end{framed}
\vspace{-10pt}
\caption{CSC pseudocode}
\label{ddos-csc}
\vspace{-10pt}
\end{figure}

\vspace{-4pt}
\noindent{\bf The contract.} The CSC contract for the above collaborative DDoS attack is outlined in Figure~\ref{ddos-csc}. We model a data feed (oracle) as a sequence of pairs $(m, \sigma)$. Where $m$ is the attack result reported by the oracle, and $\sigma$ is the oracle's digital signature over $m$. Thus, the oracle has an private/public key pair $(pk_O, sk_O)$ used to sign/verify signatures. In the figure, $bal[X]$ denotes the balance of users' account $X$ on the blockchain.

\section{Game Theoretic Model and Analysis}
\label{game-model}
\vspace{-4pt}
Dealing with self-interested attackers is problematic; they do not care about the sponsor's goals, and they would lie to collect more rewards if they can. Attacking real world targets complicates the problem since attackers will tend to hide such information to avoid any legal consequences. This raises the question of how a CSC can measure the individual contribution of each attacker. Our goal is to identify and analyze factors that influence the attacker's behavior and use that to configure incentives properly and encourage faithful collaboration. Towards this goal, in this section, we formulate our framework for CSC-based collaborative attacks as an incomplete information game, and we show that it has a strong dominant strategy equilibrium. For simplicity, we present this modeling and analysis in the context of DDoS attack, but it can be generalized to any other attack type using the proper metrics.

\vspace{-4pt}
\subsection{Attackers Contribution}
\vspace{-4pt}
Our solution utilizes betting to avoid direct measuring of individual contributions. The attack sponsor places an initial award and each attacker places a bet, all deposited in the CSC account. An attacker is supposed to contribute in the attack in proportion to his bet value. Based on how successful the attack is (as reported by the oracle), the total amount of currency in the CSC account will be distributed among the attackers and/or the sponsor (the sponsor gets a refund only if the attack is not fully successful).

Recall that all CSC information is public on the blockchain as part of the CSC code and state. However, and given that attackers are known only by their random-looking public keys, we cannot guarantee that each attacker will place only one bet. In fact, an attacker may use that to gain privacy, as well as increase his utility (if possible), by dividing the bet into multiple smaller bets, then post these bets using several accounts all controlled by this attacker. Thus, each attacker privately knows his true bet (consisted of several smaller bets), even though all the transactions are public.

Therefore, in our framework, attackers have to decide on two actions. First, before the attack phase starts, they should decide on a betting strategy: each attacker can submit multiple bets that sum to his true bet or one bet that is equal to his true bet. We call the latter \emph{truthful betting}. Afterward, in the attack phase, each attacker should decide on the amount of contribution to the attack.

Later in this section, we show that each attacker's relative contribution is independent of others' contributions and bets and is a function of his own bet value. In the next section, we also show that truthful betting is the best strategy for an attacker.

\subsection{Interdependent Attackers Game (IAG)}
\label{iag-model}
\vspace{-4pt}
We study the interaction among a set of interdependent strategic attackers as a game with independent private values and strict incomplete information. Independent private values means the utility of an attacker depends entirely on his own private information. Strict incomplete information means we have no probabilistic information in the model, i.e., we consider a worst-case scenario for missing information.

In this game, let $N$ represent the attackers set such that $|N| = n$,  $\omega_S$ be the award of the sponsor, $bet_i$ be the true bet value of the $i^{th}$ attacker such that $0 < bet_i < \omega_S$, and $t_i = \frac{bet_i}{\omega_S}$ be the private information that this attacker has and it represents his type.\footnote{We assume dealing with homogeneous agents in terms of the cost of contributing to the attack (all have same $\alpha$ in equation \ref{eq2}). Also, we show later that an attacker's bet value represents his actual contribution in the attack, and hence, his cost.}
Let $e_{th}$ represent an estimation of the total traffic needed to satisfy the desired DDoS attack result set by the sponsor, and $e_i$ be the relative contribution of the $i^{th}$ attacker in this amount, i.e., $0 \leq e_i \leq 1$. Thus, if attacker $i$ is a free-rider and makes no effort to contribute in the attack, then $e_i = 0$. While if $e_i = 1$, this means that attacker $i$ has launched a fully successful attack on his own. Let $T$ denote the set of all type profiles $\hat{t} =(t_1, \dots ,t_n)$ and $E$ denote the set of all action profile $\hat{e} = (e_1, \dots ,e_n)$. Based on his type $t_i$, attacker $i$ chooses the action $e_i \in [0,1]$. The strategy function $S: T\rightarrow E$ maps each attacker type to an action $e_i$. Rational attackers will select a strategy that will maximize their own utility as possible.

The reward allocation function is one of the mechanism rules that has a significant impact on the attacker's preferred strategies and attack result. In our framework, the sponsor defines the reward allocation function $R: [0,1] \times [0,1]  \rightarrow \mathbb{R}$ in the smart contract. We propose a simple allocation function that allots reward to attackers according to the amount of their bets and the their total relative contributions in the attack denoted as $e_{tot}$, i.e., $e_{tot} = \sum_{i\in N} e_i $.\footnote{Note that in reality we cannot compute $e_i$. Hence, $e_{tot}$ is computed by using a suitable function to convert the delay reported by the oracle into the proper total traffic relative value. For example, if the measured delay meets the desired value, i.e., fully successful attack, then $e_{tot} = 1$. If it is 50\% the value of the desired attack result, then $e_{tot} = 0.5$, and so on.} Let $M$ be the total money amount deposited in the CSC's account, so $M = \omega_S + \sum_{i\in N}bet_i$. After the attack phase ends, the CSC distributes $M$ among the attackers and the sponsor, and automatically generate payment transactions to transfer the allocated rewards to their accounts.

We consider the following reward allocation function (where $bet_{tot} = \sum\limits_{i\in N} bet_i$): 
\vspace{-4pt}
\begin{equation}
\label{reward-one}
	R(bet_i, e_{tot}) = M \cdot e_{tot} \cdot \frac{bet_i}{bet_{tot}}
\vspace{-2pt}
\end{equation}

That is, the share of each attacker from $M$ is proportional to his bet value. As noted, if $e_{tot} < 1$, meaning that the attack is not fully successful, then the residual of $M$ will go back to the sponsor. For our analysis, we want the formulation of $R$ to depend on the attacker type $t_i$. As such, equation~\ref{reward-one} can be converted into an equivalent formulation as follows: 
\vspace{-4pt}
\begin{equation}
\label{reward-two}
	R(t_i, e_{tot}) = M \cdot t_i \cdot e_{tot} \cdot \Big( \frac{bet_{tot}}{\omega_S}\Big)^{-1}
\vspace{-4pt}
\end{equation}

In order to compute the utility of an attacker, we need also to characterize the cost of preforming an attack. Let $C: [0,1] \rightarrow \mathbb{R}^+$ be a cost function that maps a value $e_i$ to the cost expended in achieving the attack contribution profile $e_i$. We assume that $C$ is strictly increasing and convex on its domain. In particular, we use the following general form cost function:
\vspace{-4pt}
\begin{equation}
\label{eq2}
	C(e_i) = \alpha \cdot \frac{\exp{(e_i)}-1}{e_{max}-e_i}\ \ \ \ \ \ \ \ \ \ \ \forall i=1, \dots,n
\vspace{-4pt}
\end{equation}

\noindent where $e_{max}$ is the max possible attack traffic (the one beyond any attacker capability), such that the cost of generating attack traffic will approach infinity when $e_i$ approaches $e_{max}$. And $\alpha$ is the average cost factor for generating attack traffic, where its value depends on the type of attack, the target's defensive power, and attackers' resources. The resulting utility function $U: [0,1] \times [0,1] \rightarrow \mathbb{R}$ for the $i^{th}$ attacker can be computed as $U\left(bet_i, e_i, e_{tot}\right) = R(t_i, e_{tot}) - C(e_i) - bet_i$, which again we convert in terms of $t_i$ as follows:
\vspace{-2pt}
\begin{equation}
\label{utility-fun}
	U\left(t_i, e_i, e_{tot}\right) = R(t_i, e_{tot}) - C(e_i) - t_i \cdot \omega_S
\end{equation}

Let $\hat{e}_{-i}$ denotes the vector components consisting of elements of $\hat{e}$ other than the $i^{th}$ element, so $e_i + \hat{e}_{-i} = e_{tot}$. Based on that, we can write the utility function as $U(t_i, e_i,\hat{e}_{-i})$, which emphasizes that the $i^{th}$ attacker only has control over his own attack effort, $e_i$. A static strict incomplete information game is then defined by a tuple $ IAG = < N,\ E,\ T,\ U > $. In the $IAG$, each attacker (aka player) maximizes its own utility in a distributed fashion. Formally, the non-cooperative $IAG$ is expressed as:
\vspace{-4pt}
\begin{equation}
\label{eq4}
\underset{e_i\in [0,1]}{\max}\ \ {U\left({t_i,e}_i,\ \hat{e}_{-i}\right)\ \ \ \forall\ i\in N}
\end{equation}

\subsection{Equilibrium Analysis}
\label{sec42}
It is necessary to characterize a set of attack efforts where each player is satisfied with the utility he receives (which maximizes his utility), given the attack efforts of other players. Such an operating point is called an equilibrium. The equilibrium concept offers a predictable, stable outcome of a game where multiple players with conflicting interests compete through self-optimization and reach a point where no player wishes to deviate. The condition for stability we aim for here is that of strong dominant strategy equilibrium. Note that rules of the game, including the utilities of all players, are public knowledge but not their private information, namely, $t_i$ and $e_i$.

First, we derive the best-response strategy of a player in $IAG$. Then we prove that the described game among attackers in this setting has a dominant strategy equilibrium. The $i^{th}$ attacker's best response strategy $S^\ast(t_i, \hat{e}_{-i})$ to a given strategy profile $\hat{e}_{-i}$ is given as the unconstrained maximizer of his utility, where $AW$, $bet_{tot}$ and $e_{th}$ are fixed:
\vspace{-2pt}
\begin{equation}
\label{eq5}
S^\ast(t_i, \hat{e}_{-i})=\underset{e_i\in [0,1]}{\arg\max}\ {U\left({t_i,e}_i,\ \hat{e}_{-i}\right)}
\vspace{-2pt}
\end{equation}

To find the maximizing $S^\ast(t_i)$, we take the first derivative of $U$ with respect to $e_i$ and equate it to 0:
\vspace{-2pt}
\begin{equation}
\label{eq6}
-\alpha \cdot \frac{\exp{(e_i)}}{e_{max}-e_i}-c\cdot\frac{\exp{(e_i)}-1}{(e_{max}-e_i)^2}+\frac{\omega_S \cdot t_i \cdot (\omega_S + bet_{tot})}{bet_{tot}}=0
\end{equation}
\vspace{-2pt}

So, the best response strategy $S^\ast(t_i, \hat{e}_{-i})$ of the $i^{th}$ player (with a value denoted as $e_i^\ast$) does not depend on the attack efforts of the other players $\hat{e}_{-i}$. Therefore, we can represent the best response strategy function by $S^\ast\left(t_i\right)$. Furthermore, the only parameters (other than $ t_i $) that determine $S^\ast\left(t_i\right)=e_i^\ast$ are the cost of the required attack traffic $\alpha$ and the quantity $\frac{bet_{tot}}{\omega_S}$.

\begin{definition}
Given an incomplete information game $\Gamma =<N, E, T, U>$, a strategy $S^\ast(t_i)$ is a \textit{strongly dominant strategy}, if for every $ t_i $ we have that the strategy $S^\ast(t_i)$ is a strongly dominant strategy in the full information game defined by $ t_i $. Formally, for all $t_i$, all $\hat{e}_{-i}$, and all possible values for $e_i$ (denoted as $e_i^\prime$) such that $e_i^\prime \neq S^\ast(t_i)$, we have
\vspace{-2pt}
\begin{equation}
\label{eq7}
U(t_i, S^\ast(t_i), \hat{e}_{-i}) > U(t_i, e_i^\prime,\hat{e}_{-i})
\vspace{-2pt}
\end{equation}
\end{definition}

\begin{definition}
Given an incomplete information game $\Gamma =<N, E, T, U>$, a strong dominant strategy equilibrium is an action profile $\hat{e}^\ast = (S^\ast(t_1 ), \dots, S^\ast (t_n ))$ in which each $S^\ast(t_i)$ is a strongly dominant strategy. The notion of strongly dominant strategy requires that $S^\ast(t_i)$ is the unique best response to all possible $\hat{e}_{-i}$, i.e., without knowing anything about $\hat{t}_{-i}$ (the type of other attackers).
\end{definition}

\begin{theorem}\label{thm1}
$IAG$ defined above has a strong dominant strategy equilibrium.
\end{theorem}
\begin{proof}
For each player $i$ and for fixed $t_i$ and $e_{-i}$, the reward function $R: T \times E\rightarrow \mathbb{R}$ is linear, and cost function $C: E\rightarrow \mathbb{R}$ is convex with respect to $e_i$. So $U(t_i, e_i,\hat{e}_{-i})$ is concave and has a unique maximum.
$S^\ast(t_i)$ is the unique maximizer of $U(t_i, e_i,\hat{e}_{-i})$\footnote{Note that the utility function is increasing at $e_i=0$ and decreasing at $e_i=1$, so the maximum can not occur at end points} and is a strongly dominant strategy that is the best response regardless of $\hat{e}_{-i}$. So for any $i \in N$ and all $t_i$, all $\hat{e}_{-i}$ and all $e_i^\prime$, we have $U(t_i, S^\ast(t_i), \hat{e}_{-i}) > U(t_i, e_i^\prime, \hat{e}_{-i})$ and strategy profile $\hat{e} = (S^\ast(t_1), \dots, S^\ast(t_n))$ is the strong dominant strategy equilibrium of the game. \qed	
\end{proof}

\section{Exploring Incentive Compatibility}
\label{inc-comp}
As discussed earlier, due to the anonymity characteristic of blockchain users, we can not determine the number of attackers $n$ and the amount of their individual bets $bet_i$. In fact, knowing these bets allows predicting the attack result in the equilibrium of the game, i.e., $\sum_{i\in N} S^\ast(t_i) = e_{tot}^\ast$. In this section, we model the rules that govern the interactions in the CSC as a mechanism. By applying mechanism design theory, we prove that despite private information and pure selfish behavior, we can predict the attack result and the conditions that impact it. In addition, we show that under certain reasonable conditions, our mechanism satisfies the necessary constraints of mechanism design, namely, incentive compatibility, individual rationality, budget balance, and fairness.

\vspace{-4pt}
\subsection{Mechanism Formulation}
\vspace{-4pt}
As we showed in the previous section, the $IAG$ game modeling players' interactions in a CSC has a strong dominant strategy equilibrium. As such, in equilibrium, the contribution of the $i^{th}$ attacker with type $t_i$ will be $S^\ast(t_i) = e_i^\ast$, and it is independent of other attackers' types and strategies.  Based on that, for an attacker type profile $\hat{t} \in T$, we define the attack result function $AR: T \rightarrow E$ as:
\vspace{-2pt}
\begin{equation}
\label{eq8}
AR(\hat{t}) = \sum_{i\in N}{S^\ast(t_i)} = \sum_{i\in N} e_i^\ast = e_{tot}^\ast
\vspace{-2pt}
\end{equation}

We consider the $i^{th}$ attacker's bet $bet_i$ as his payment to CSC, and the reward $R(t_i,AR(\hat{t}))$ is what the mechanism pays him. Accordingly, this attacker receives a payment amount $p_i$, or makes it if $p_i$ is negative, expressed as (note that $t_i \cdot \omega_S = bet_i$): 
\vspace{-2pt}
\begin{equation}
\label{eq9}
p_i(\hat{t}) = R(t_i,AR(\hat{t})) - t_i \cdot \omega_S
\vspace{-2pt}
\end{equation}

Let $G(\hat{t}) = (S(\hat{t}),P(\hat{t}))$ be the outcome function that maps each type profile $\hat{t} \in T$ to an outcome $o = (e_{tot}^\ast,\hat{p})$ where the payment rule $P(.)$ defines a profile of attackers' payments $\hat{p} = ({p_0,p}_1, \dots, p_n)$, and the set of possible outcomes is denoted by $O$.

The attack result $e_{tot}^\ast$ is the non-monetary part in the outcome. To monetize it, we introduce a valuation function $V: \mathbb{R}^+ \times T \rightarrow \mathbb{R}$ (in terms of some currency) to represent an attackers' preference for a given attack result. In other words, the valuation function expresses the cost that an attacker is willing to tolerate when contributing to the attack (based on his dominant strategy), in addition to the transaction fee paid to post his bets on the CSC (i.e., fees for Ethereum miners). So, if this attacker post his true bet in $k$ small bets, each of which will require $\delta$ transaction fee (since each will be sent in a separate transaction), the total fee will be $k \cdot \delta$. Based on that, $V$ can be expressed as follows:
\vspace{-2pt}
\begin{equation}
\label{eq10}
V(e_{tot}^\star, t_i) = - (C(S^\star(t_i)) + k \cdot \delta),\ \ \ \forall i\in N, \forall t_i \in [0,1]
\vspace{-2pt}
\end{equation}

Recall that attacker $i$'s contribution depends only on his type $t_i$, and so is the valuation function shown in equation~\ref{eq10}. Thus, we can express it as $V(e_{tot}^\star, t_i) = V(t_i) $. As a result, this valuation function can be used to represent the type of attacker, and announcing a type is similar to reporting the attacker's valuation function.

Since we have money/incentive transfer between agents (aka attackers), we work in quasi-linear setting. Each agent has a utility that is the motivating factor behind the selection of his strategy. The preference of attacker $i$ can be captured using his utility function that can be redefined as follows (this is equivalent to the one defined in equation~\ref{utility-fun}):
\vspace{-4pt}
\begin{equation}
\label{eq11}
U(t_i, o) = V(t_i ) + p_i
\vspace{-4pt}
\end{equation}

We formalize the incentive mechanism for strategic attackers in the proposed collaborative attack as a direct mechanism.\footnote{This mechanism suits our model since we have the space of possible actions is equal to the space of possible types, so an attacker type (which is defined when he bets) is the same as his act (the amount of attack contribution).} In this setup, each attacker $i$ announces a type $t_i^\prime$ to the mechanism, which is not necessarily equal to his true type $t_i \in [0,1] $, such that it will lead to an outcome that maximizes his utility. We also have a social choice function $F: T \rightarrow O$ that maps each agents' type profile to an optimal outcome, which is the same as the outcome function.

\begin{definition}
A direct mechanism in a quasi-linear setting is defined by $D = (T, G(\hat{t}))$. The mechanism defines the set of allowable types $T$ that each agent can choose and an outcome function $G$ which specifies an outcome $o$ for each possible type profile $\hat{t} = (t_1, \dots, t_n) \in T$.
\end{definition}

\subsection{Incentive Compatible Property}
\vspace{-4pt}
Direct mechanisms extract information from agents by motivating them to \emph{tell the truth}. If the best response for all attackers to report their private information truthfully to the CSC-driven mechanism, we say the contract is incentive compatible. Here we prove that the proposed mechanism is cheat-proof, which means that all attackers are motivated to submit their bet truthfully, and any deviation will lead to a utility loss.

\begin{definition}
The social choice function $F(\cdot)$ is a dominant strategy incentive compatible (DSIC) (aka strategy-proof or cheat-proof) if and only if
\vspace{-4pt}
\begin{equation}
\label{eq12}
U(t_i, o)\geq\ U(t_i^\prime, o^\prime)\ \ \ \   \forall i\in N, \ \forall\hat{t}\in T, \ \forall t_i^\prime\in [0,1]
\vspace{-4pt}
\end{equation}

\noindent where $o=G(t_i,\hat{t}_{-i})$ and $o^\prime = G(t_i^\prime,\hat{t}_{-i})$.
\end{definition}

Thus, if the SCF is DSIC, then the best response for agent $i$ is to bet truthfully, i.e., $bet_i = t_i \cdot \omega_S$, regardless of other attackers' bets. By calling a direct mechanism DSIC or strategy-proof, we mean that the mechanism implements an incentive-compatible or strategy-proof social choice function.

\begin{theorem}
\label{thm2}
The proposed direct mechanism (or social choice function) modeling our CSC-based collaborative attacks is DSIC.
\end{theorem}

\begin{proof}
We need to show that the utility of attacker $i$, is maximized when he bet truthfully. We use proof by contradiction. Without loss of generality, assume that the true type of attacker $i$ is $t_i$, but he tries to misrepresent his type and submit his bet with two distinct blockchain addresses in two transactions ($k=2$) containing two bets $bet_{i,1}$ and $bet_{i,2}$, such that $bet_{i,1}+ bet_{i,2} = bet_i$ and $bet_i = t_i\cdot \omega_S$. We denote the incorrect types by $t_{i,1}$ and $t_{i,2}$, so we have $t_i = t_{i,1} + t_{i,2}$. We assume that he benefits from this untruthful act which means $\exists t_{i,1}, t_{i,2} \in T$, where $t_i = t_{i,1} + t_{i,2}$, and $\exists\hat{t}_{-i}\in T^{n-1}$ such that 
\vspace{-4pt}	
\begin{equation}
\label{eq13}
U(o^\prime, t_{i,1}) + U(o^\prime, t_{i,2}) > U(o, t_i)
\vspace{-4pt}
\end{equation}
where $G(t_{i,1}+t_{i,2},\hat{t}_{-i})$ and $G(t_i,\hat{t}_{-i})$ are the values of $o^\prime$ and $o$, respectively.
	
Note that, from each attacker $ i $'s perspective, his dominant strategy $S^\ast(t_i)$ is a function of his true type $t_i$, which he knows, and so he can calculate his dominant strategy as $S^\ast(t_i)=S^\ast(t_{i,1}+t_{i,2})$. Thus: 
	
\vspace{-4pt}
\begin{align}
\label{eq14}
AR(t_{i,1}+t_{i,2}, \hat{t}_{-i}) &= S^\ast(t_{i,1}+t_{i,2})+ \sum_{j \in N, i \ne j}S^\ast (t_j) \\
&= S^\ast(t_i) \;\;\;\;\;\;\;\;\;\;\;\; + \sum_{j \in N,i\ne j}S^\ast (t_j) = AR(t_i,\hat{t}_{-i}) \notag
\vspace{-4pt}
\end{align}

And for the rewards, we have:
\vspace{-4pt}
\begin{align}
\label{eq15}
R(t_{i,1}, e_{tot}) + R(t_{i,2}, e_{tot}) 
&= (t_{i,1} + t_{i,2}) \cdot (bet_{tot} + \omega_S) \cdot \Big(\frac{bet_{tot}}{\omega_S} \Big)^{-1} \cdot e_{tot} \\
\vspace{-2pt}
&= t_i\cdot (bet_{tot} + \omega_S) \cdot \Big(\frac{bet_{tot}}{\omega_S} \Big)^{-1} \cdot e_{tot} = R(t_i,e_{tot}) \notag 
\vspace{-4pt}
\end{align}

So the attacker's reward does not increase by hiding his type (i.e., hiding his bet). Therefore, from the mechanism point of view, the payments are the same as if the attacker would announce his true type, so: 
\vspace{-2pt}
\begin{equation}
\label{eq16}
p_i(t_i,\hat{t}_{-i}) = p_i(t_{i,1}, \hat{t}_{-i}) + p_i(t_{i,2},\hat{t}_{-i})
\vspace{-2pt}
\end{equation}

As we see, rewards and payments are identical in both cases. So the only way for an attacker to obtain a higher utility when lying, is to have a higher valuation than the one for the betting truthfully. This means that we have:
\vspace{-4pt}
\begin{equation}
\label{eq17}
V(t_{i,1})+V(t_{i,2}) > V(t_i)
\vspace{-4pt}
\end{equation}
	
Which means that:
\vspace{-4pt}
\begin{equation}
\label{eq18}
C(S^\ast(t_i))+ 2\delta < C(S^\ast(t_i))+\delta
\vspace{-4pt}
\end{equation}
This inequality cannot hold since this means that a transaction fee $\delta$ is negative, which is not the case. Thus, we a get a contradiction, meaning that an attacker will not get a higher utility by submitting multiple bets instead of the a single one (i.e., his true bet). \qed
\end{proof}

Note that we assumed attackers do not trust each other, so each attacker at least needs one transaction to submit his bet or he can submit multiple bets that sum to his true bet. However, let’s assume that two attackers do trust each other and want to collude and fool the mechanism to increase their profit. The only possible misbehavior is to submit one bet that its amount is equal to the sum of their bets. In this case, the mechanism considers them as one identity (i.e., one attacker) and based on the proposed reward allocation function, this will not increase their reward. The only effect of this collusion is that the actual attack result would be higher than the pre-calculated attack result. This is because the same amount of contribution costs less for two attackers than one attacker (recall that the cost function is convex) and when two attackers assumed as one, their contribution will be considered less.

\subsection{Budget Constraint}
To be economically feasible, an incentive mechanism must be budget constrained. In our framework, the total rewards allocated to the attackers should not exceed $M = \sum_{i\in N}bet_i + \omega_S$, which is the total deposit made to the CSC account. In terms of payments, this means that the total attackers payment should not exceed the sponsor award $\omega_S$.

\begin{definition}
(Budget Constraint for CSC). A reward mechanism is budget constrained if for $\forall \hat{t} \in T$ we have $\sum_{i\in N}{p_i(\hat{t})} \le \omega_S$.
\end{definition}

\begin{theorem}
	\label{thm3}
	The proposed direct mechanism is budget constrained. 
\end{theorem}

\begin{proof}
The total attackers payments can be expressed as:
\vspace{-2pt}
\begin{align}
\label{eq20}
\sum_{i\in N}{p_i(\hat{t})} &= \sum_{i\in N}\Big(R(t_i, S(\hat{t}))-t_i \cdot \omega_S\Big) \notag \\ \vspace{-2pt}
&=\sum_{i\in N}\Big((bet_{tot}+ \omega_S) \cdot t_i \cdot \Big(\frac{bet_{tot}}{\omega_S}\Big)^{-1} \cdot S(\hat{t})-t_i \cdot \omega_S\Big) \\ \vspace{-2pt}
&= (bet_{tot} + \omega_S) \cdot AR(\hat{t})-bet_{tot} \notag
\vspace{-2pt}
\end{align}
 
\noindent where $AR:T\rightarrow E$ is as given in equation \ref{eq8} and $E=[0,1]$. So for any type profile $\hat{t}\in T$ we have:
\begin{equation}
\label{eq21}
-bet_{tot} \le (bet_{tot} + \omega_S) \cdot S(\hat{t})-bet_{tot} \le \omega_S
\end{equation}
	
\noindent From equations~\ref{eq20} and~\ref{eq21}, we get $\sum_{i\in N}{p_i(\hat{t})\le \omega_S}$, completing the proof. \qed
\end{proof}

\subsection{Voluntary Participation Constraint}
\label{vol-part}
Individual rationality or voluntary participation property of a social choice function means that each attacker gains a non-negative utility by participating in the mechanism that implements the social choice function. There are two stages at which individual rationality can be examined. First, when the amount of bet (types) of other attackers $\hat{t}_{-i}$ is unknown to attacker $ i $, and therefore predicting attack result (i.e. outcome) is impossible. Second, when before the attack phase (ex-post stage), a choice to withdraw from the mechanism is given to all attackers. That is when all the attackers have announced their bet, and an attack result can be calculated. Note that a truthful attacker who submits his bet in one transaction incurs two transaction fees, $2\delta$, to withdraw from the mechanism (and if an attacker submits $k$ smaller bets, his cost to withdraw will $2k\delta$). This property of\textit{ ex-post individual rationality} is stated as follows.

\begin{definition}
(Ex-post Individual Rationality). The utility an attacker $i$ with type $t_i$ receives by withdrawing from a CSC is equal to $-2 k \delta$. To ensure attacker $i$'s participation when withdrawal is allowed at the ex-post stage, we must satisfy the following ex-post Individual Rationality (or participation) constraint
\vspace{-4pt}
\begin{equation}
\label{eq22}
U(G(\hat{t}),t_i)\geq -2 k \delta\ \ \ \ \forall\hat{t}\in T
\vspace{-4pt}
\end{equation}
\end{definition}

A mechanism satisfies ex-post individual rationality if it implements a social choice function that satisfies ex-post individual rationality. In the next section, through numerical analysis, we show that under mild conditions the proposed mechanism satisfies this constraint. Note that these conditions can be encoded as rules in CSC by the sponsor. For instance, he can condition the attack on a specified amount of total bets, or he can restrict attackers' type by specifying upper bound and lower bound on acceptable bets.

\subsection{Fairness}
Different definitions for fairness have been proposed in the literature including proportional fairness, max-min fairness, $ \alpha- $fairness \cite{trichakis2011fairness}. To measure the fairness of the proposed incentive mechanism, we need a metric that captures how close the payment obtained by this mechanism is to fair payment. By fair payment, we mean reward allocation based on the contribution of attackers (instead of the amount of their bets). In other words, if attackers could prove their contribution, then the reward allocation function would simply be a function of their contribution and the attack result. 

We denote the fair payment by $p_i'$ and can be computed as:
\vspace{-2pt}
\begin{equation}
\label{eq23}
p_i'(\hat e) = \omega_S \cdot e_i \cdot e_{tot}
\vspace{-2pt}
\end{equation}

Similar to \cite{da2009resource}, we consider a metric, which we call the fairness-score, based on the root mean square (RMS) of the difference between the mechanism payment and the fair payment. The fairness-score is defined as
\vspace{-2pt}
\begin{equation}
\label{eq24}
D_{rms} = \sqrt {\frac{1}{n} \sum_{i\in N} ( p_i(\hat t) - p_i' (\hat e))^2}
\vspace{-2pt}
\end{equation}

\noindent where $ p_i(\hat t) $ is given by equation \ref{eq9}. A value of zero for $ D_{rms} $ indicates that the mechanism payment equals to the fair payment, which discourages free-riders and rewards contributors. As the value increases, the fairness of the reward allocation function decreases. In the next section, using numerical simulation for two reward functions, we show that under some mild conditions, the fairness score drops for both rewarding schemes after hitting the desired attack result. \vspace{-8pt}

\section{Numerical Simulations and Discussion}
\label{analysis}
Based on our game model, there are four key parameters that impact the attack outcome (in terms of payment and attack result). These factors are $bet_{tot}$, $\omega_S$, $C(e_{th})$, and type profile $\hat{t} = (t_1, \dots, t_n)$. Furthermore, the reward allocation function has a significant impact on the attack result and other properties, such as fairness and individual rationality.

In this section, we conduct simulations for a typical DDoS scenario to analyze how various system parameters impact attack outcome. The parameter values used in this study are $n=30$, $\frac{bet_{max}}{bet_{min}}=\frac{1}{10}$, $bet_i \sim Uniform(bet_{min},bet_{max})$, and the results are averaged over 50 randomly chosen attackers' bets. In our simulations, we consider two reward allocation functions: linear and quadratic (in the bet amount). The former is the one given by equation~\ref{reward-two} and denoted by $R$, while the latter is expressed as:\footnote{Note that the theoretical (game and mechanism) analysis conducted for $R$ holds for $R'$, too.} 
\vspace{-2pt}
\begin{equation}
\label{eq25}
R'(\hat t, e_{tot})= M \cdot e_{tot} \cdot \frac{t_i^2}{\sum_{i\in N}t_i^2}
\vspace{-2pt}
\end{equation}

As for the attack result, it will have an upper bound based on the contract terms. That is, putting more effort beyond what is needed to reach the desired attack result will not increase the rewards for the attackers. Thus, rational attacker will stop when the measured attack result reaches the desired value.

We define two parameters: $\theta = \frac{bet_{tot}}{\omega_S}$, and $\gamma = \frac{C(e_{th})}{\omega_S}$. Here,  $\theta$ gives an indication of the total amount of bets with respect to the sponsor award, and $\gamma$ is an indicator of the cost of performing the attack with respect to this award. We use these parameters in our simulations to study the impact of total bet and cost on the attack result.

\begin{figure}[t!]
\centering
\begin{subfigure}[b]{.32\textwidth}
	\centering
	\includegraphics[width=\textwidth]{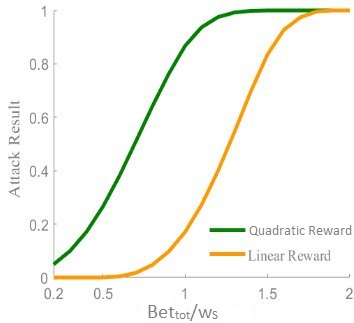}
	\caption{\hspace{1cm}}
\end{subfigure}
\begin{subfigure}[b]{.32\textwidth}
	\centering
	\includegraphics[width=\textwidth]{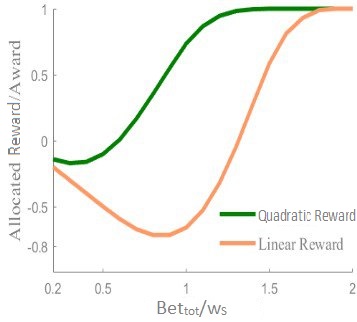}
	\caption{\hspace{1cm}}
\end{subfigure}
\begin{subfigure}[b]{.32\textwidth}
	\centering
	\includegraphics[width=\textwidth]{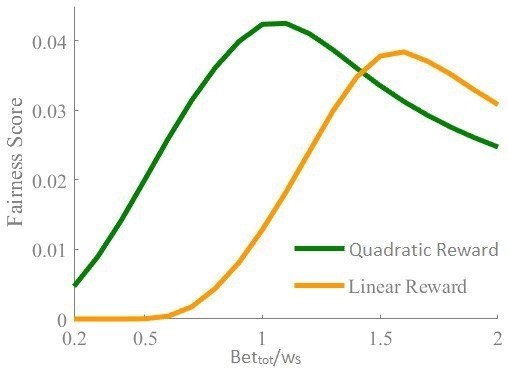}
	\caption{\hspace{1cm}}
\end{subfigure}
\vspace{-8pt}
\caption{The impact of increasing $\theta$ for fixed $\gamma = 0.35$ on (a) attack result (b) proportion of allocated award to attackers, and (c) fairness-score of payment.}
\label{Fig456}
\vspace{-12pt}
\end{figure}

Comparing the impact of reward allocation functions, as Figure~\ref{Fig456}(a) shows the quadratic scheme attracts significantly more contribution than the linear reward scheme with the same value of $\theta$. In this scheme, having $\theta = 1$, one can estimate that the expected attack result $e_{tot}^\ast$ would be around $90\%$ of the desired attack result $e_{th}$. On the other hand, we observe that the linear reward allocation function takes $\theta$ $\approx$ $1.7$ to attract such contribution however this scheme allocate the rewards fairer compared to the quadratic scheme. Moreover, Figure~\ref{Fig456}(b) shows that at that point the quadratic reward scheme pays $75\%$ of the considered award while it costs attackers around $80\%$ of the total cost of a successful attack. The remaining $25\%$ of the award will be paid for reaching the desired attack result $e_{th}$. As expected, the attack result, the allocated reward to award ratio, and fairness score increase with the increase of $\theta$. Therefore, the attack sponsor can condition the attack on a minimum amount of total bet to ensure the success of the attack and a given fairness score. Also, Figure~\ref{Fig456}(c) demonstrates that after hitting the desired attack result, in both rewarding schemes, the fairness-score drops which means after this point, the more the total bet, the fairer the allocation will be.

\begin{figure}[t!]
	\centering
	\begin{subfigure}[b]{.32\textwidth}
		\centering
		\includegraphics[width=\textwidth]{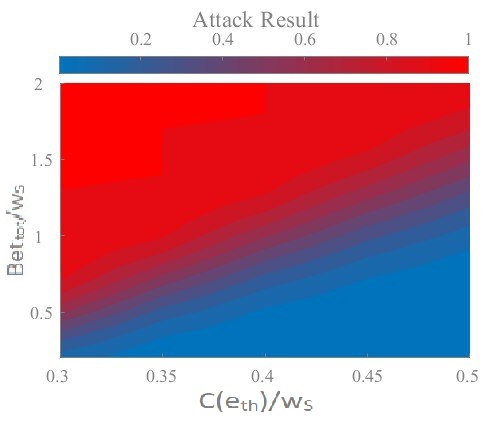}
		\caption{\hspace{1cm}}
	\end{subfigure}
	\begin{subfigure}[b]{.32\textwidth}
		\centering
		\includegraphics[width=\textwidth]{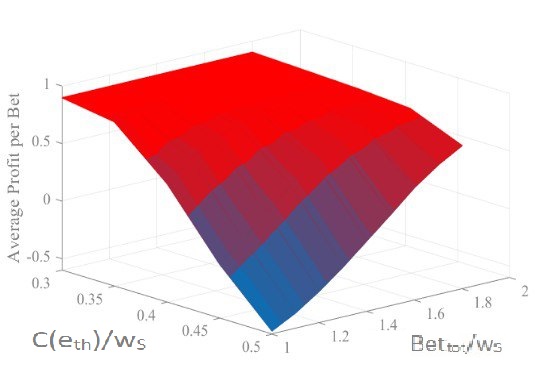}
		\caption{\hspace{1cm}}
	\end{subfigure}
	\begin{subfigure}[b]{.32\textwidth}
		\centering
		\includegraphics[width=\textwidth]{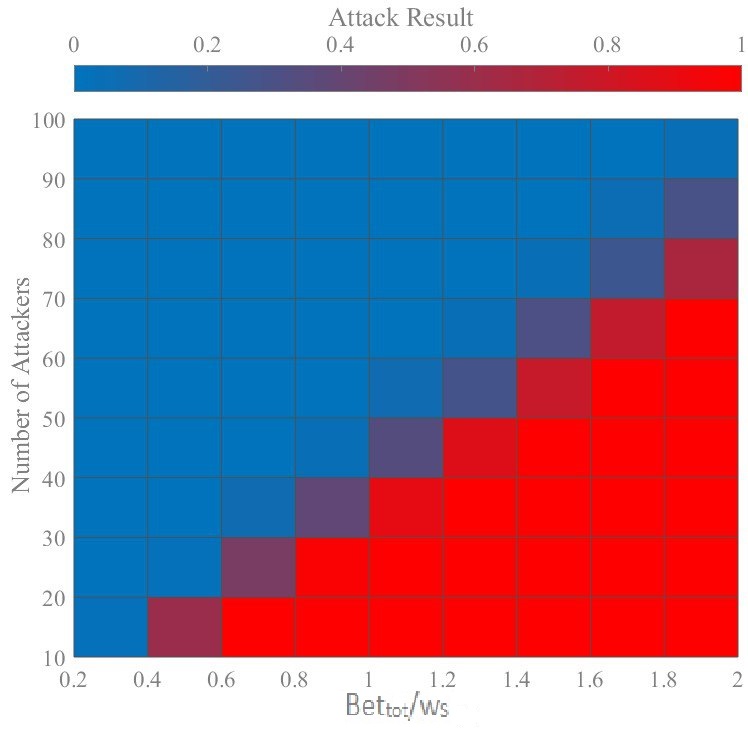}
		\caption{\hspace{1cm}}
	\end{subfigure}
	\vspace{-2pt}
	\caption{The impact of $ \theta $ and $ \gamma $ on (a) the attack-result, and (b) ratio of average profit over average bet size, with fixed number of attackers $ n = 30 $. (c) the impact of $\theta$ and number of attackers on attack result ($\gamma=0.35$)}
	\label{Fig89}
	\vspace{-4pt}
\end{figure}

As Figure~\ref{Fig89}(a) shows, the required $\theta$ for providing the incentive of launching a successful attack increases as the cost factor $\gamma$ increases. Therefore, as the attack cost goes higher, a larger value of $bet_{tot}$ is needed to provide the required incentives to launch a successful attack. Accordingly, knowing the total amount of the bets, the number of attackers, and the award, the target server can estimate and change the attack result through increasing the cost of that attack (be deploying proper defenses against DDoS).

As discussed in section~\ref{vol-part}, attackers are incentivized to participate when a non-negative utility is expected. Figure~\ref{Fig89}(b) shows that under some mild conditions, in terms of $\theta$ and $\gamma$, the proposed mechanism satisfies the desirable property of individual rationality. The red region denotes the situations that the average attackers' profit is positive. That is knowing the cost of the desired attack and the minimum total bet, the attack sponsor or the attack target can adjust the award or the cost to incentivize or deincentivize attackers to participate.

As shown in Figure~\ref{Fig89}(c), the larger the number of attackers the larger $bet_{tot}$ required for launching a successful attack. In other words, increasing the number of attackers alone would not lead to more collaboration rather they should be incentivized enough to collaborate. Therefore accepting small bets decreases the attack result and a rule of minimum acceptable bet in the CSC can be helpful.

\section{Conclusion}
\label{conclusion}
In this paper, we introduced a framework for employing CSCs to orchestrate real-world targeted attacks. These attacks are launched by several collaborating attackers without any knowledge of each other or any trust between them. To study the feasibility of this idea, we considered DDoS attack as a usecase. By using thorough game theoretic analysis and mechanism design, we showed that the attack sponsor can design a cheat-proof and budget-balanced mechanism to encourage collaboration of selfish rational attackers. Furthermore, the sponsor can predict and adapt the attack result, i.e., determine under what conditions attackers will participate in the attack. Simulation results show that, under some mild conditions on the attack cost and total amount of bets, the proposed incentive mechanism provides individual rationality and fair allocation of rewards. Being the first to study CSC-based collaborative attacks against real-work targets, we believe that our work will contribute in promoting the foundational understanding of these attacks, an important step towards developing effective countermeasures.


\bibliographystyle{splncs03}
\bibliography{paper}


\end{document}